\documentclass[3p,authoryear]{elsarticle}
\usepackage{amsmath,amsfonts,amssymb,amsthm}
\usepackage{psfrag}
\newtheorem{definition}{Definition}
\newtheorem{theorem}[definition]{Theorem}
\def\rz{{\mathbb{R}}}
\def\nz{{\mathbb{N}}}
\begin{document}

\title{Monte Carlo methods in statistical physics:\\ mathematical foundations and strategies}

\author{Michael Kastner\corref{cor1}}
\ead{kastner@sun.ac.za}
\address{National Institute for Theoretical Physics (NITheP), Stellenbosch 7600, South Africa}

\begin{abstract}
Monte Carlo is a versatile and frequently used tool in statistical physics and beyond. Correspondingly, the number of algorithms and variants reported in the literature is vast, and an overview is not easy to achieve. In this pedagogical review, we start by presenting the probabilistic concepts which are at the basis of the Monte Carlo method. From these concepts the relevant free parameters---which still may be adjusted---are identified. Having identified these parameters, most of the tangled mass of methods and algorithms in statistical physics Monte Carlo can be regarded as realizations of merely a handful of basic strategies which are employed in order to improve convergence of a Monte Carlo computation. Once the notations introduced are available, many of the most widely used Monte Carlo methods and algorithms can be formulated in a few lines. In such a formulation, the core ideas are exposed and possible generalizations of the methods are less obscured by the details of a particular algorithm.
\end{abstract}

\maketitle


\section*{Preface}

This text provides a gentle and mostly self-contained introduction to Monte Carlo methods in statistical physics. It focuses on the mathematical and foundational aspects of Monte Carlo, showing under which conditions a Monte Carlo calculation can be expected to converge, and what parameters may be tuned in order to improve convergence.

The presentation of the collected material differs in style and rigour from what is typically found in physics textbooks. It should be useful as supplementary material for an undergraduate or graduate course, but also scientists getting started on Monte Carlo-related research should profit from its reading. Finally, even the experienced Monte Carlo practitioner should find some less-known background material of interest in the present article, or he might just enjoy the more abstract point of view onto the subject. In particular, a glossary is provided which translates some of the most widely used algorithms into the formal language developed in the main body of the article.

\section{Introduction}

The vast majority of interacting many-particle systems in physics defies an analytic solution. Therefore it is no surprise that numerical methods have become increasingly popular over the last decades, up to the point where computational physics, besides experimental and theoretical physics, is considered as one of the three major pillars of physics. Monte Carlo methods are among the most widely and frequently used numerical techniques, with applications in statistical physics, quantum mechanics, field theory, and others. Throughout the present article, the language of Monte Carlo methods in classical statistical physics is used, but the concepts presented are not restricted to that field.

In its simplest version, a Monte Carlo simulation consists of only a few lines of computer code, and this simplicity renders it relatively easy to get started on the subject. Therefore it is no surprise that virtually any textbook on the subject starts by describing something like the ``standard version'' of a Monte Carlo simulation (single spin flip Metropolis algorithm) and then proceeds to more advanced algorithms by modifying some ingredients of this standard version [see \cite{LandauBinder} and \cite{NewmanBarkema} for introductory textbooks on Monte Carlo simulations in statistical physics]. When learning about advanced methods, it becomes obvious that the general Monte Carlo scheme has several ``free parameters'' which may be adjusted in order to improve performance. A clear identification of these free parameters is often missing, and the range in which they can be varied is given only implicitly. Even among scientists employing Monte Carlo methods, these issues are not as well-known as one might suppose. As I found out in many discussions, there is still an astonishingly large number of people believing that, in order to compute a Monte Carlo estimator of, say, a canonical expectation value, one necessarily has to perform a canonical simulation with Boltzmann-type acceptance rates. This is not only incorrect, but it proved an obstacle for the invention of new and more efficient Monte Carlo algorithms in the past.


A formal presentation of the foundations of Monte Carlo---and that is the main point to be made in this article---is useful in order to clearly identify the remaining free parameters of the underlying Markov chain. The choice of these parameters is at the basis of both, the power and the flexibility of the Monte Carlo method, and following a few basic strategies, convergence of the Monte Carlo simulation may be improved significantly. This knowledge is essential to fully exploit the possibilities Monte Carlo has to offer and to creatively develop new, efficient algorithms.

Though only little of the material presented here is new, a good fraction of it has not been cast into the present form before. In other instances, certain results are well-known in the mathematical literature, but not so among statistical physicists (Theorem \ref{is_theorem} being an example). Other topics, like random number generation, error estimation, and the more practical aspects of Monte Carlo are not covered in this article.

In Sec.\ \ref{markov}, the few basic facts from the theory of stochastic processes which are required for Monte Carlo are reviewed, allowing us to introduce Markov-chain Monte Carlo in Sec.\ \ref{mcsim}. A formal description of how to construct transition matrices of a Markov chain is then presented in Sec.\ \ref{constr}. From this formalization, it becomes evident which parameters may be adjusted in order to improve convergence. In Sec.\ \ref{design}, the strategies to improve the performance of a Monte Carlo simulation are introduced. Methods and strategies going beyond Markov-chain Monte Carlo are briefly discussed in Sec.\ \ref{summary}, together with a summary of the results presented. The paper is supplemented by a glossary, listing some of the most frequently used Monte Carlo methods in the language put forward in the main body of the article. 


\section{Markov chains}
\label{markov}

The characteristic feature of Monte Carlo methods is that stochasticity, i.e.\ randomness, is used to obtain an approximation of a quantity of interest. A Markov chain is the type of stochastic process which is typically employed and to which we will restrict ourselves throughout this paper. Understanding the notion of a Markov chain will be almost all which is necessary for the definition of Monte Carlo.\footnote{Note that, for brevity, the term ``Monte Carlo'' is used in the following, although ``Markov chain Monte Carlo'' would be more precise.}

First we will define the ``arena'' on which a Markov chain is to be constructed.
\begin{definition}[State space]
Let\/ $S$ be a set with a finite number of elements. Each element of\/ $S$ is called a\/ {\em state}, and\/ $S$ is called the\/ {\em state space}.
\end{definition}
In classical statistical physics, the state space is typically the {\em configurational space}\/ or the {\em phase space}\/ of the system. Which state space we have to choose will become obvious from what is called a ``quantity of interest'' in Definition \ref{def_mcestimator}.

The restriction of finiteness is imposed on $S$ merely for simplicity, and a generalization of the following to a state space with an infinite number of elements is possible. However, regarding the finiteness of the computer on which we intend to implement the Monte Carlo method later on, we will be confined to a finite state space anyway. If the state space of the system we want to investigate is infinite, for example in the case of continuously varying values of positions and momenta or continuously varying spin variables, we will be forced to do some sort of discretization or coarse-graining in order to obtain a finite state space, suitable for an implementation on a computer.

Although finite, the number of states will in general be very large. It is this fact which forces us to revert to an approximate analysis like the Monte Carlo method instead of an exact calculation. In very many applications in statistical physics, the state space has a product structure, being the product of the state spaces of the (large number of) subsystems constituting the statistical system. This product structure is often exploited when constructing the {\em proposal matrix} (see Sec.\ \ref{constr} and several entries in the Glossary).

In order to describe the stochasticity involved in Markov chains, a few elementary definitions from probability theory are needed.
\begin{definition}[Distribution]
A vector\/ $\lambda=(\lambda_i:i\in S)$ is called a distribution on $S$ if its elements\/ $\lambda_i$ satisfy the conditions
\begin{enumerate}
\item $\lambda_i\geq0$ for all\/ $i\in S$,
\item $\sum\limits_{i\in S} \lambda_i=1$.
\end{enumerate} 
\end{definition}
We will use distributions to describe the probabilities of the random outcome of the Monte Carlo method.
\begin{definition}[Stochastic matrix]\label{def_stoch}
A matrix\/ ${\mathtt T}=({\mathtt T}_{ij}:i,j\in S)$ is called a stochastic matrix on $S$ if all its row vectors are distributions, i.e.
\begin{enumerate}
\item ${\mathtt T}_{ij}\geq0$ for all\/ $i,j\in S$,
\item $\sum\limits_{i\in S} {\mathtt T}_{ij}=1$ for all\/ $j\in S$.
\end{enumerate} 
\end{definition}
It is important to note (and easy to verify) that the stochasticity property of a matrix ${\mathtt T}$ ensures that, if $\lambda$ is a distribution, the product ${\mathtt T}\lambda$ yields again a distribution. Thus, from a given distribution, we can generate a set of distributions by multiplication(s) with a stochastic matrix.

For a given matrix ${\mathtt T}$, there might exist one or more distributions which show the special property of {\em stationarity} with respect to ${\mathtt T}$.
\begin{definition}[Stationary distribution]
A stationary distribution of a matrix\/ ${\mathtt T}$ is a distribution\/ $\pi$ such that\/ ${\mathtt T}\pi=\pi$.
\end{definition}
Rephrased in terms of linear algebra, $\pi$ is a right eigenvector of ${\mathtt T}$ with eigenvalue $1$. These stationary distributions will play an important role for the question whether, and at which speed, a Monte Carlo estimator converges towards the quantity of interest (addressed in Sec.~\ref{mcsim}).
\begin{definition}[Random variable]
A random variable\/ $X$ on a finite state space\/ $S$ is a variable which takes on values from the state space\/ $S$, depending on the outcome of a random experiment. If the probability to obtain a certain value\/ $i\in S$ as an outcome of the experiment is\/ $\lambda_i$, the vector\/ $\lambda=(\lambda_i:i\in S)$ is said to be the distribution of\/ $X$.
\end{definition}
For our purposes this will do as a working definition of a random variable. For a rigorous definition of a random variable as a mapping from a probability space to a measure space, see for example the textbook see \citet{Bauer}. As a simple example, the reader might consider a random variable which describes the outcome of the toss of a coin, where the state space $S=\{\text{head},\text{tail}\}$ consists of two possible outcomes of the experiment with probabilities $\lambda_\text{head}=\lambda_\text{tail}=1/2$.

Now we will gather the preceding definitions to define a Markov chain as a certain sequence of random variables.
\begin{definition}[Markov chain]\label{def_markov}
Let\/ ${\mathtt T}$ be a stochastic matrix and\/ $X_n$ a random variable on\/ $S$. The sequence\/ $\left\{X_n\right\}_{n=0}^{N-1}$ of random variables\/ $X_n$ is called a Markov chain of length\/ $N$ on\/ $S$ with transition matrix ${\mathtt T}$ and initial distribution $\lambda$ if
\begin{enumerate}
\item $X_0$ has distribution\/ $\lambda$,
\item for\/ $n\geq 1$, conditional on\/ $X_n=i$, the random variable\/ $X_{n+1}$ has distribution\/ $({\mathtt T}_{ij}:j\in S)$.
\end{enumerate}
Such a chain is called $\mbox{Markov}(\lambda,{\mathtt T},N)$.
\end{definition}
It is item two of this definition which distinguishes Markov chains from other stochastic processes: the fact that the distribution of the random variable $X_{n+1}$ depends exclusively on its immediate predecessor $X_n$, but not on earlier ones like $X_{n-1},X_{n-2},...$ ``A Markov chain has no memory'' is an expression frequently used to put this property into words. The Markov property is convenient for the implementation of a stochastic process on a computer: One is not obliged to keep track of the history of the system, as only the current state of the system has to be considered in order to specify the next.

The transition matrix ${\mathtt T}$ contains all the probabilities necessary to obtain $X_{n+1}$ from its predecessor $X_n$. If we have $X_n=j$, the probability to find $X_{n+1}=i$ is ${\mathtt T}_{ij}$. As a simple consequence it is found that, if $X_0=j$, the probability to find $X_n=i$ is given by $({\mathtt T}^n)_{ij}$, the $ij$-th element of the $n$-th power of the transition matrix ${\mathtt T}$ \citep[see][for a proof]{Norris}. We can therefore write a $\mbox{Markov}(\lambda,{\mathtt T},N)$ chain explicitly as the sequence
\begin{equation}
\left\{X_n\right\}_{n=0}^{N-1}=\left\{\lambda,{\mathtt T}\lambda,{\mathtt T}^2\lambda,\dots,{\mathtt T}^{N-1}\lambda\right\}.
\end{equation}

With the notion of a Markov chain at hand, we are now in the position to define the Monte Carlo method.

\section{Monte Carlo}
\label{mcsim}

In equilibrium statistical physics, one is often interested in quantities which involve a summation or integration over the state space of a system. Typical examples include partition functions or ensemble averages of quantities like energy or magnetization. Even for systems consisting of a modest number of particles, the state space can be very large, too large in fact to perform such a summation or integration even on a computer.

The fundamental concept of the Monte Carlo method is to replace the summation over the state space $S$ by a summation over a Markov chain on $S$. Thus, instead of summing over all elements from $S$, only a sample of randomly chosen states is considered. This will of course lead to an estimator of the desired quantity instead of an exact result.
\begin{definition}[Monte Carlo estimator]\label{def_mcestimator}
Let $F=\sum\limits_{i\in S}f(i)$ be a quantity of interest, where $f$ is a function on $S$. We call
\begin{equation}
\widetilde{F}_{\lambda,{\mathtt T},N}=\frac{1}{N} \sum\limits_{i\in\{X_n\}_{n=0}^{N-1}} \frac{f(i)}{\pi_i}
\end{equation}
the Monte Carlo estimator of $F$, where $\{X_n\}_{n=0}^{N-1}$ is $\mbox{Markov}(\lambda,{\mathtt T},N)$ on $S$, and $\pi=(\pi_i:i\in S)$ with $\pi_i>0$ is the stationary distribution of the transition matrix ${\mathtt T}$.
\end{definition}
As an example, consider as a quantity of interest the canonical expectation value of the energy,
\begin{equation}\label{eq:Ecan}
E=\langle H\rangle = \frac{1}{Z(\beta)}\sum_{i\in S}H(i)\exp[-\beta H(i)],
\end{equation}
where $H$ is the Hamiltonian, $\beta$ is the inverse temperature $1/T$ (with Boltzmann's constant set to unity), and
\begin{equation}
Z(\beta)=\sum_{i\in S}\exp[-\beta H(i)]
\end{equation}
is the canonical partition function. Then we define $f(i)=H(i)\exp[-\beta H(i)]/Z(\beta)$. If, by chance or by habit, we happen to choose a Markov chain with
\begin{equation}\label{eq:Bdistr}
\pi=(\pi_i=\exp[-\beta H(i)]/Z(\beta):i\in S)
\end{equation}
as its stationary distribution, we find that the Monte Carlo estimator of the canonical expectation value of the energy,
\begin{equation}
\widetilde{E}_{\lambda,{\mathtt T},N} = \frac{1}{N} \sum\limits_{i\in\{X_n\}_{n=0}^{N-1}} H(i),
\end{equation}
is simply given by the average of $H(i)$ over the Markov chain.

Note that, although it might appear ``natural,'' it is by no means necessary to employ a Markov chain with a Boltzmann-type stationary distribution \eqref{eq:Bdistr}. If, for whatsoever reason, we had preferred a Markov chain with, say, a Gaussian stationary distribution
\begin{equation}
\pi=(\pi_i\propto\exp\{-a[H(i)-b]^2\}:i\in S)
\end{equation}
with parameters $a$ and $b$, we still could have computed a Monte Carlo estimator of the canonical expectation value of the energy by means of the formula
\begin{equation}
\widetilde{E}_{\lambda,{\mathtt T},N} = \frac{\sum_{i\in\{X_n\}_{n=0}^{N-1}} H(i)\exp[-\beta H(i)]/\pi_i}{\sum_{i\in\{X_n\}_{n=0}^{N-1}} \exp[-\beta H(i)]/\pi_i}.
\end{equation}
Although the Boltzmann distribution is part of the given physical problem, it need not be part of the problem's solution. In fact, it will depend on the physical system of interest whether it is a good idea or not to use the Boltzmann distribution also as the stationary distribution of the Markov chain. We will come back to this issue in Sec.\ \ref{design}.

The rest of this paper (like virtually all scientific work published on Monte Carlo) is dedicated to the question of how to optimize the quality of this estimator. As indicated by the subscripts of $\widetilde{F}$ in Definition \ref{def_mcestimator}, the Monte Carlo estimator depends crucially on the parameters of the Markov chain employed. These parameters are therefore at our disposal in order to tune the Monte Carlo method:
\begin{enumerate}
\item The initial distribution $\lambda$,
\item the transition matrix ${\mathtt T}$, and
\item the length $N$ of the Markov chain (number of elements).
\end{enumerate}
It is in particular the transition matrix which is modified in order to increase the quality of the Monte Carlo estimator. However, this is done under the constraint of computational efficiency, meaning that a too complicated choice of ${\mathtt T}$ would, with the same amount of CPU time on a computer, lead to a significantly lower length $N$ of the Markov chain which can be achieved. The task of constructing a transition matrix appropriate for the system under investigation and the quantity of interest is by no means trivial, and Sec.\ \ref{design} will be devoted to some aspects of this issue.

Up to now, we do not even know whether, and under which conditions, we have a fair chance of obtaining a useful estimator of the quantity of interest at all. In the remainder of this section, the notion of convergence is defined for a Monte Carlo estimator, and a condition on the transition matrix is derived which ensures convergence. Convergence, albeit not a guaranty, constitutes something like a minimal requirement for a Monte Carlo simulation to give reasonable results.

\begin{definition}[Convergence of a Monte Carlo estimator]\label{def_conv}
A Monte Car\-lo estimator\/ $\widetilde{F}_{\lambda,{\mathtt T},N}$ of\/ $F$ is said to converge if, for a given transition matrix\/ ${\mathtt T}$,
\begin{equation}
\lim_{N\to\infty}\widetilde{F}_{\lambda,{\mathtt T},N}=F
\end{equation}
almost surely for all initial distributions\/ $\lambda$.
\end{definition}
When convergence is guaranteed, we know that, at least in the limit of infinite length of the Markov chain, the method provides a Monte Carlo estimator arbitrarily close to the exact result of the quantity of interest.

In Definition \ref{def_conv}, the remaining free parameter is the transition matrix ${\mathtt T}$. To specify the conditions on the transition matrix which ensure convergence, we need the notions of {\em aperiodicity}\/ and {\em irreducibility}.
\begin{definition}[Aperiodic matrix]
A matrix ${\mathtt T}=({\mathtt T}_{ij}:i,j\in S)$ is called aperiodic if, for any $i\in S$, there exists a natural number $n_{i}\in\nz$ such that $({\mathtt T}^{n})_{ii}>0$ for all $n\geqslant n_i$.
\end{definition}
\begin{definition}[Irreducible matrix]
A matrix ${\mathtt T}=({\mathtt T}_{ij}:i,j\in S)$ is called irreducible if, for any $i,j\in S$, there exists a natural number $n_{ij}\in\nz$ such that $({\mathtt T}^{n_{ij}})_{ij}>0$.
\end{definition}
As already mentioned, the $n$-th power of the transition matrix ${\mathtt T}$ yields the distribution of the $n$-th element of the Markov chain, conditional on the initial distribution. Hence, in case the transition matrix is aperiodic and irreducible, for {\em all}\/ sufficiently large $n$ we have a non-zero probability to find {\em any} state of the state space, independently of the initial distribution. As a consequence, an aperiodic and irreducible transition matrix prevents us from ``getting stuck'' in any partition of the state space while ``leaving out'' another part. This is an essential prerequisite of the following theorem.
\begin{theorem}[Ergodic theorem]
Let\/ ${\mathtt T}$ be an aperiodic and irreducible stochastic matrix on a finite state space\/ $S$. Then there exists a unique stationary distribution\/ $(\pi_i:i\in S)$ of\/ ${\mathtt T}$. Further, let\/ $\{X_n\}_{n=0}^{N-1}$ be\/ $\mbox{Markov}(\lambda,{\mathtt T},N)$. Then, for a bounded function\/ $f:S\to\rz$,
\begin{equation}
\lim_{N\to\infty}\frac{1}{N}\sum_{i\in\{X_n\}_{n=0}^{N-1}}f(i)=\sum_{i\in S}\pi_i f(i)
\end{equation}
almost surely for any initial distribution\/ $\lambda$. 
\end{theorem}
{\em Proof:} See for example \citet{Norris}, Ch.\ 1.10.

The ergodic theorem allows us to read off immediately the conditions on a Monte Carlo estimator guaranteeing convergence: Employing a Markov chain with an aperiodic and irreducible transition matrix, a Monte Carlo estimator converges in the limit of infinite chain length.

In an implementation of the Monte Carlo method on a computer, however, the length of the Markov chain is related to the running time on a computer, and is therefore finite. Hence, for a given system, we are left with the question which choices of transition matrices yield good estimators within a certain finite time. Before addressing this problem in Sec.~\ref{design}, we will describe a method how to construct a transition matrix with a given stationary distribution.

\section{Construction of transition matrices}
\label{constr}

We have seen that a stationary distribution is a characteristic feature of a transition matrix. In applications, one typically has in mind a certain distribution $\pi$, like the canonical one in Eq.\ \eqref{eq:Bdistr}, and the goal is to construct a transition matrix having that $\pi$ is its stationary distribution. As pioneered by \citet{MeRoRoTeTe}, this is conveniently done by constructing the transition probabilities from a proposal matrix ${\mathtt P}=({\mathtt P}_{ij}:i,j\in S)$ and an acceptance matrix ${\mathtt A}=({\mathtt A}_{ij}:i,j\in S)$. Starting from the element $X_n=j$ of a Markov chain, a {\em proposal}\/ is made according to the proposal probabilities ${\mathtt P}_{ij}$ for $i$ to be the next element of the Markov chain. The proposal is {\em accepted}, setting $X_{n+1}=i$, with probability ${\mathtt A}_{ij}$. It is rejected, setting $X_{n+1}=j$, with probability $(1-{\mathtt A}_{ij})$. Formalizing this procedure, the elements of the transition matrix ${\mathtt T}$ can be written as
\begin{equation}\label{TausPA}
{\mathtt T}_{ij}={\mathtt P}_{ij}{\mathtt A}_{ij}+\delta_{ij}\sum_{k\in S}{\mathtt P}_{kj}(1-{\mathtt A}_{kj})
\end{equation}
where $\delta$ denotes Kronecker's delta symbol ($\delta_{ij}=1$ if\/ $i=j$, $\delta_{ij}=0$ else).

Why is this commonly done in applications of the Monte Carlo method? First, such a procedure is straightforward to implement on a computer. Second, as will be discussed below, sufficient conditions on ${\mathtt P}$ and ${\mathtt A}$ can be specified which guarantee a certain distribution to be the stationary distribution of the resulting transition matrix. To describe these conditions, the notion of {\em detailed balance} is introduced.%
\footnote{An algorithm for the construction of a transition matrix with a given stationary distribution, but {\em without} the condition of detailed balance for the acceptance matrix has been described by \citet{Boghosian}. 
}
\begin{definition}[Detailed balance]
Let\/ ${\mathtt A}=({\mathtt A}_{ij}:i,j\in S)$ be a matrix and\/ $\pi=(\pi_i:i\in S)$ be a vector. ${\mathtt A}$ is said to satisfy detailed balance with respect to\/ $\pi$ if
\begin{equation}
{\mathtt A}_{ij}\pi_j={\mathtt A}_{ji}\pi_i
\end{equation}
for all\/ $i,j\in S$.
\end{definition}
Now we are ready to give a recipe of how to construct a transition matrix with a given stationary distribution.
\begin{theorem}[Transition matrix with given stationary distribution]\label{tm_theorem}
Let\/ ${\mathtt P}=({\mathtt P}_{ij}:i,j\in S)$ and\/ ${\mathtt A}=({\mathtt A}_{ij}:i,j\in S)$ be matrices and\/ $\pi=(\pi_i:i\in S)$ be a distribution subject to the following conditions:
\renewcommand{\labelenumi}{C\arabic{enumi}.}
\begin{enumerate}
\item \label{Psym}${\mathtt P}$ is symmetric, i.\,e., ${\mathtt P}_{ij}={\mathtt P}_{ji}$ for all\/ $i,j\in S$,
\item \label{Pstoch}${\mathtt P}$ is stochastic (see Definition \ref{def_stoch}),
\item \label{Apos}$0\leq {\mathtt A}_{ij} \leq 1$ for all\/ $i,j\in S$,
\item \label{Adetbal}${\mathtt A}$ satisfies detailed balance with respect to\/ $\pi$.
\end{enumerate}
Then the transition matrix\/ ${\mathtt T}$ as defined in Eq.~(\ref{TausPA}) is a stochastic matrix with stationary distribution\/ $\pi$.
\end{theorem}
{\em Proof:} see Appendix \ref{tm_proof}.

\citet{Hastings} gives a similar recipe for the construction of transition matrices from non-symmetric proposal matrices. In order to construct an acceptance or transition matrix which satisfies detailed balance, a simple recipe given by \citet{CaPeSo} can be employed.
\begin{theorem}[Matrix which satisfies detailed balance]\label{acc_detbal}
Let\/ $f:(0,\infty)\to[0,1]$ be a function satisfying $f(z)=z\,f(1/z)$ for all\/ $z$. Then the matrix\/ ${\mathtt A}=({\mathtt A}_{ij}:i,j\in S)$ with elements\/ ${\mathtt A}_{ij}=f(\pi_i/\pi_j)$ satisfies detailed balance with respect to\/ $\pi=(\pi_i:i\in S)$.
\end{theorem}
{\em Proof:}
\begin{equation}
{\mathtt A}_{ij}\pi_j=f\left(\frac{\pi_i}{\pi_j}\right)\pi_j=\frac{\pi_i}{\pi_j}f\left(\frac{\pi_j}{\pi_i}\right)\pi_j=f\left(\frac{\pi_j}{\pi_i}\right)\pi_i={\mathtt A}_{ji}\pi_i.
\end{equation}
Note that, when constructing the acceptance matrix ${\mathtt A}$ by this recipe, only the ratios $\pi_i/\pi_j$ of the elements of the stationary distribution $\pi$ enter. This is an important feature since this way the normalization constant of $\pi$ is not required (and computing this normalization would again involve a summation over the entire state space). It is for this reason that in the following (and in particular in the glossary) prominent examples of distributions used for Monte Carlo computations are defined as proportionalities.

The most common choice for the {\em acceptance function} in Theorem \ref{acc_detbal} is $f:z\mapsto\min\{1,z\}$, associated with the names of \citet{MeRoRoTeTe}. An alternative is the Glauber acceptance function $f:z\mapsto z/(1+z)$.

So what have we gained so far? We have described an algorithm which tells us how to construct a transition matrix with a given stationary distribution. Although from this algorithm only a subset of all transition matrices can be constructed, there are still plenty of choices to be made within this subset, namely:
\renewcommand{\labelenumi}{P\arabic{enumi}.}
\begin{enumerate}
\item An initial distribution,
\item a symmetric, stochastic proposal matrix,
\item a stationary distribution, and
\item an acceptance function.
\end{enumerate}
\renewcommand{\labelenumi}{\arabic{enumi}.}

The vast majority of Monte Carlo calculations applied in statistical mechanics use transition matrices which fit into this scheme of ``proposal and acceptance,'' and prominent choices of proposal matrices, stationary distributions, and acceptance functions are summarized in the Glossary at the end of this paper.

Being free to choose a distribution $\pi$ on a state space $S$ and then construct a transition matrix ${\mathtt T}$ such that $\pi$ is the stationary distribution of ${\mathtt T}$, we have to cope with the fact that the number of elements in the state space is typically very large. The only manageable procedure is to define $\pi$ by means of a---preferably simple---functional dependence of its elements,
\begin{equation}\label{pig}
\pi=(\pi_i=g(i):i\in S),\quad\mbox{where}\quad g:S\to[0,1].
\end{equation}
In statistical physics there exist one or more functions on phase space or configurational space which play a fundamental role for the system under investigation. The most prominent example is the Hamiltonian function, but one might also think of the magnetization of a spin system, a sub-lattice magnetization, and many more. It appears to be ``natural,'' mainly due to the concept of {\em importance sampling} described in Sec.\ \ref{importance}, to employ these function(s) also for the construction of the transition matrix. This is typically done by choosing a stationary distribution expressed in terms of, for instance, the Hamiltonian function ${\mathcal H}$, and the standard example is clearly the canonical (Boltzmann) distribution $\pi_i\propto\exp[-\beta{\mathcal H}(i)]$, with inverse temperature $\beta$. Note that properties of the physical system under investigation enter into the transition matrix, and thus into the Markov process, only via such a choice of the stationary distribution. 

As an example we now construct the transition matrix for a system of two Ising spins, using some standard choices of the Markov chain parameters. Consider a system of two degrees of freedom $\sigma_1$ and $\sigma_2$, each of which can take on values from $\{-1,+1\}$. The composite system has a product structure, i.e.\ a state $\sigma_1 \times \sigma_2$ of the composed system is an element of the product space $C=\{-1,+1\}\times\{-1,+1\}$. This is the so-called {\em configurational space} of the system which we choose as the state space of the Markov chain. The Hamiltonian function of the system, a mapping from the configurational space onto the reals, is given by
\begin{equation}
{\mathcal H}(\sigma_1,\sigma_2)=\sigma_1 \sigma_2.
\end{equation}


For notational simplicity, the four elements of $C$ are numbered as follows: $1=(+1)\times(+1)$, $2=(+1)\times(-1)$, $3=(-1)\times(+1)$, $4=(-1)\times(-1)$. The element ${\mathtt T}_{21}$ of the transition matrix, for example, yields the probability to go from the state $(+1)\times(+1)$ to the state $(+1)\times(-1)$.

As a proposal matrix, we choose a single spin flip algorithm, proposing to change any one of the two spins, each with probability 1/2:
\begin{equation}
{\mathtt P}=\frac{1}{2}\begin{pmatrix} 0&1&1&0\\ 1&0&0&1\\ 1&0&0&1\\ 0&1&1&0 \end{pmatrix}.
\end{equation}
As in the example of section \ref{mcsim}, we choose the Boltzmann distribution $\pi_i\propto {\mathrm e}^{-\beta{\mathcal H}(i)}$ to be the stationary distribution of the Markov chain,
\begin{equation}\label{eq:pi_choice}
\pi\propto\begin{pmatrix}{\mathrm e}^{+\beta}\\ {\mathrm e}^{-\beta}\\ {\mathrm e}^{-\beta}\\ {\mathrm e}^{+\beta} \end{pmatrix}.
\end{equation}
The standard Metropolis acceptance function $f(z)=\min\{1,z\}$ then leads to an acceptance matrix of the form
\begin{equation}
{\mathtt A}=\begin{pmatrix}1&1&1&1\\ {\mathrm e}^{-2\beta}&1&1&{\mathrm e}^{-2\beta}\\ {\mathrm e}^{-2\beta}&1&1&{\mathrm e}^{-2\beta}\\ 1&1&1&1 \end{pmatrix}.
\end{equation}
For the transition matrix of the Markov chain we therefore obtain
\begin{equation}
{\mathtt T}=\frac{1}{2}\begin{pmatrix}2(1-{\mathrm e}^{-2\beta})&1&1&0\\ {\mathrm e}^{-2\beta}&0&0&{\mathrm e}^{-2\beta}\\ {\mathrm e}^{-2\beta}&0&0&{\mathrm e}^{-2\beta}\\ 0&1&1&2(1-{\mathrm e}^{-2\beta}) \end{pmatrix},
\end{equation}
and it is easily verified that $\pi$ as given in Eq.\ \eqref{eq:pi_choice} is indeed its stationary distribution.

When implementing the Monte Carlo method on a computer, one would, however, never bother to compute all of the entries of the transition matrix ${\mathtt T}$. Fortunately, it is sufficient to propose a new state and compute only the acceptance rate of this particular proposal.



\section{Tuning parameters}\label{design}

In the previous sections, we have provided a basis for addressing a question which, for all practical applications of the Monte Carlo method, is of utmost importance and, at the same time, very difficult to answer in general: Given a state space and a certain quantity of interest; for which choices of the remaining parameters (initial distribution, proposal matrix, stationary distribution, and acceptance function) can we obtain a good Monte Carlo estimator of the quantity of interest within a certain amount of time? An additional complication arises from the fact that optimization is desired not with respect to Monte Carlo time (length $N$ of the Markov chain), but with respect to the CPU time of the computer on which the Monte Carlo run is implemented. Therefore, whenever possible, computationally simple choices of proposal and acceptance probabilities are to be preferred. Although this issue will not be discussed in great detail, we will come back to it later on.

It can be regarded as a general rule that, the more about the quantity of interest is already known, the better the optimization problem can be approached. {\em In praxi}, a rather limited number of strategies how to appropriately choose proposal and acceptance matrices is employed, the two most prominent ones being:
\begin{description}
\item[Importance sampling:] The stochasticity of the Mar\-kov chain gives rise to a statistical error in the Monte Carlo estimator $\widetilde{F}_{\lambda,{\mathtt T},N}$ of the quantity of interest $F$. Choose the transition matrix ${\mathtt T}$ such that this statistical error is minimized.
\item[Avoiding effective non-ergodicity:] Even if a transition matrix is ergodic, transitions between one block of states and another one may be extremely rare on the time scale accessible by computer simulation. This situation leads to a systematic error in Monte Carlo estimators, which is to be avoided.
\end{description}
In short, importance sampling is about reducing the statistical error, avoiding block-like transition matrices refers to avoiding systematic errors. In order to obtain a useful Monte Carlo estimator, both these issues have to be taken into account simultaneously (at least up to a satisfactory level), which is often a very hard task. In the following, for each of these two concepts, the above developed notations allow us to present the important and prevalent Monte Carlo strategies applied in statistical physics (but not only there) in a terse and precise way.

A third concept to improve results from Monte Carlo is by sophisticatedly choosing the quantity $F$ for which a Monte Carlo estimator is computed. Examples of such quantities include Monte Carlo renormalization \citep{Swendsen:79} and transition matrix Monte Carlo \citep{WaSwe}. These choices depend strongly on the physical problem under investigation, and a discussion of this topic is beyond the scope of this article.

\subsection{Importance sampling}\label{importance}

The aim of importance sampling is to choose the transition matrix of a Markov chain such that the statistical error present in a Monte Carlo estimator is minimal. An answer to the question of which transition matrix to choose is provided by the following theorem \citep{Rubinstein}.
\begin{theorem}[Importance sampling]\label{is_theorem}
For a set of random variables\/ $\{X_n\}_{n=0}^{N-1}$ distributed according to\/ $\pi=(\pi_i:i\in S)$, the variance of
\begin{equation}
\widetilde{F}=\frac{1}{N}\sum\limits_{i\in\{X_n\}_{n=0}^{N-1}}\frac{f(i)}{\pi_i}
\end{equation}
is minimal if the proportionality
\begin{equation}
\pi_i\propto |f(i)|\neq0
\end{equation}
holds.\footnote{The condition $|f(i)|\neq0$ is no serious restriction on the choice of functions $f$, as the states $i$ for which $f(i)=0$ could be simply excluded from the state space. Then, however, ergodicity on this reduced state space needs to be fulfilled.}
\end{theorem}
{\em Proof:} By making use of Jensen's inequality; see Appendix \ref{is_proof}.

Frequently, from a single Markov chain, several (or even a continuum of) estimators for different quantities of interest are to be computed. Examples include the simultaneous computation of different canonical averages like energy and magnetization, or the computation of a certain canonical average for various values of the temperature. In this case, what is an optimal choice for one single quantity is a suboptimal choice for another, and a compromise has to be found. The ideal compromise can be found by slightly generalizing the above theorem, yielding
\begin{equation}
\pi_i\propto\sqrt{\sum_{m=1}^{M}f_m(i)^2}
\end{equation}
as the optimal stationary distribution for the simultaneous computation of the $M$ Monte Carlo estimators 
\begin{equation}
\widetilde{F}_m=\frac{1}{N}\sum\limits_{i\in\{X_n\}_{n=0}^{N-1}}\frac{f_m(i)}{\pi_i},\qquad m=1,\dotsc,M.
\end{equation}

For the computation of a Monte Carlo estimator of, say, a canonical expectation value of the energy as defined in \eqref{eq:Ecan}, the ideal choice, following Theorem \ref{is_theorem}, of the stationary distribution of the Markov chain is 
\begin{equation}
\pi_i=|H(i)|\exp[-\beta H(i)].
\end{equation}
At first sight, it may come as a surprise to the reader that, nonetheless, such an error minimizing distribution is virtually never used in applications. Among the reasons for this are the following:
\begin{itemize}
\item Even among expert computational physicists, Theorem \ref{is_theorem} is not very well-known.
\item If, for a certain choice of the stationary distribution, the computation of the transition probabilities ${\mathtt T}_{ij}$ is costly in time, it may pay off to choose a 'similar' but computationally simpler stationary distribution. This allows to obtain a longer Markov chain within the same amount of CPU time and may thus reduce the statistical error. An example is, again, the Boltzmann distribution, where the transition probabilities depend only on the energy {\em differences}, not on their absolute values. Especially for systems with short-range interactions this fact can be exploited to gain computational advantages.
\item There are situations where, in spite of Theorem \ref{is_theorem}, the quality of the Monte Carlo estimator obtained with a different stationary distribution may be superior: The choice of the stationary distribution also has an effect on whether or not the transition matrix has a block-like structure (as discussed in the next section), possibly leading to systematic errors in the Monte Carlo estimators. It may pay off to choose a stationary distribution which is not optimal in the sense of importance sampling, but which avoids a block-like transition matrix. An example is the umbrella sampling (or multicanonical or entropic sampling) method for systems whose infinite counterparts show a first order phase transition (see Glossary and, for example, \citet{BeNe}).
\end{itemize}

\subsection{Avoiding effective non-ergodicity}\label{speed}

An often encountered problem which prevents convergence of a Monte Carlo estimator on the time scale accessible is the presence of an almost block-like structure in powers ${\mathtt T}^n$ of the transition matrix for large $n$. ``Almost block-like'' refers to a situation where, for a suitable choice of the basis, all elements of ${\mathtt T}^n$ in the exterior of some blocks on the diagonal are exceedingly small. This situation can be sketched as ${\mathtt T}^n$ being of the form
\[
{\mathtt T}^n = \begin{pmatrix}
\parbox{1cm}{\includegraphics[width=1cm,height=1cm]{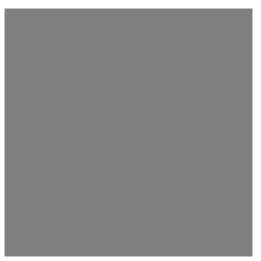}}
\parbox{1cm}{\center $\gtrapprox 0$\vphantom{$\int\limits_m$}}\\
\parbox{1cm}{\center $\gtrapprox 0$\vphantom{$\int\limits_m$}}
\parbox{1cm}{\includegraphics[width=1cm,height=1cm]{matrix.eps}}
\end{pmatrix}
\]
for all sufficiently large $n$. Although, in a strict sense, ${\mathtt T}$ might be ergodic, this leads to ``effective non-ergodicity,'' meaning that for viable values of the Markov chain length $N$, certain parts of the state space will not be reached from a given initial distribution with very high probability. Hence, the actual realization of the Markov chain differs significantly from its expected distribution in the limit $N\to\infty$, leading to a bad Monte Carlo estimator in general. For a transition matrix constructed according to the recipe from Sec.\ \ref{constr}, the emergence of a block-like structure will originate from the interplay of the chosen proposal matrices ${\mathtt P}$ and acceptance matrices ${\mathtt A}$. Since usually, when constructing acceptance matrices in statistical physics, the Hamiltonian or related phase space functions enter, occurrence of a block-like transition matrix will depend strongly on the physical system under investigation. The presence of a first-order phase transition is an example where such problems are encountered frequently. Due to the dependence on the properties of the physical system considered, it is difficult to devise a general solution. Instead, we will briefly outline the two principle strategies of addressing this problem by means of an example.

Consider the Ising model on a two-dimensional square lattice, characterized by the Hamiltonian
\begin{equation}
H(\sigma)=-J\sum_{\langle i,j\rangle}\sigma_i \sigma_j,
\end{equation}
where $J>0$ is a coupling constant and the angular brackets denote a summation over all pairs of neighboring sites on the lattice. Each of the $\sigma_i$ ($i=1,\dotsc,N$) is a classical spin variable, taking on the values $-1$ or $+1$. The states $\sigma=(\sigma_1,\dotsc,\sigma_N)$ are therefore elements from the product space $S=\{-1,+1\}^N$.

\begin{figure}[hb]\center
	\psfrag{ 0}{{\scriptsize $0$}}
	\psfrag{ 0.2}{{\scriptsize $0.2$}}
	\psfrag{ 0.4}{{\scriptsize $0.4$}}
	\psfrag{ 0.6}{{\scriptsize $0.6$}}
	\psfrag{ 0.8}{{\scriptsize $0.8$}}
	\psfrag{ 1}{{\scriptsize $1$}}
	\psfrag{ 1.2}{{\scriptsize $1.2$}}
	\psfrag{-1}{{\scriptsize $-1$}}
	\psfrag{-0.5}{{\scriptsize $-0.5$}}
	\psfrag{ 0.5}{{\scriptsize $0.5$}}
	\psfrag{m}{{\small $m$}}
	\psfrag{P}{{\small $\mathcal{P}$}}
	\includegraphics[scale=0.48,angle=270]{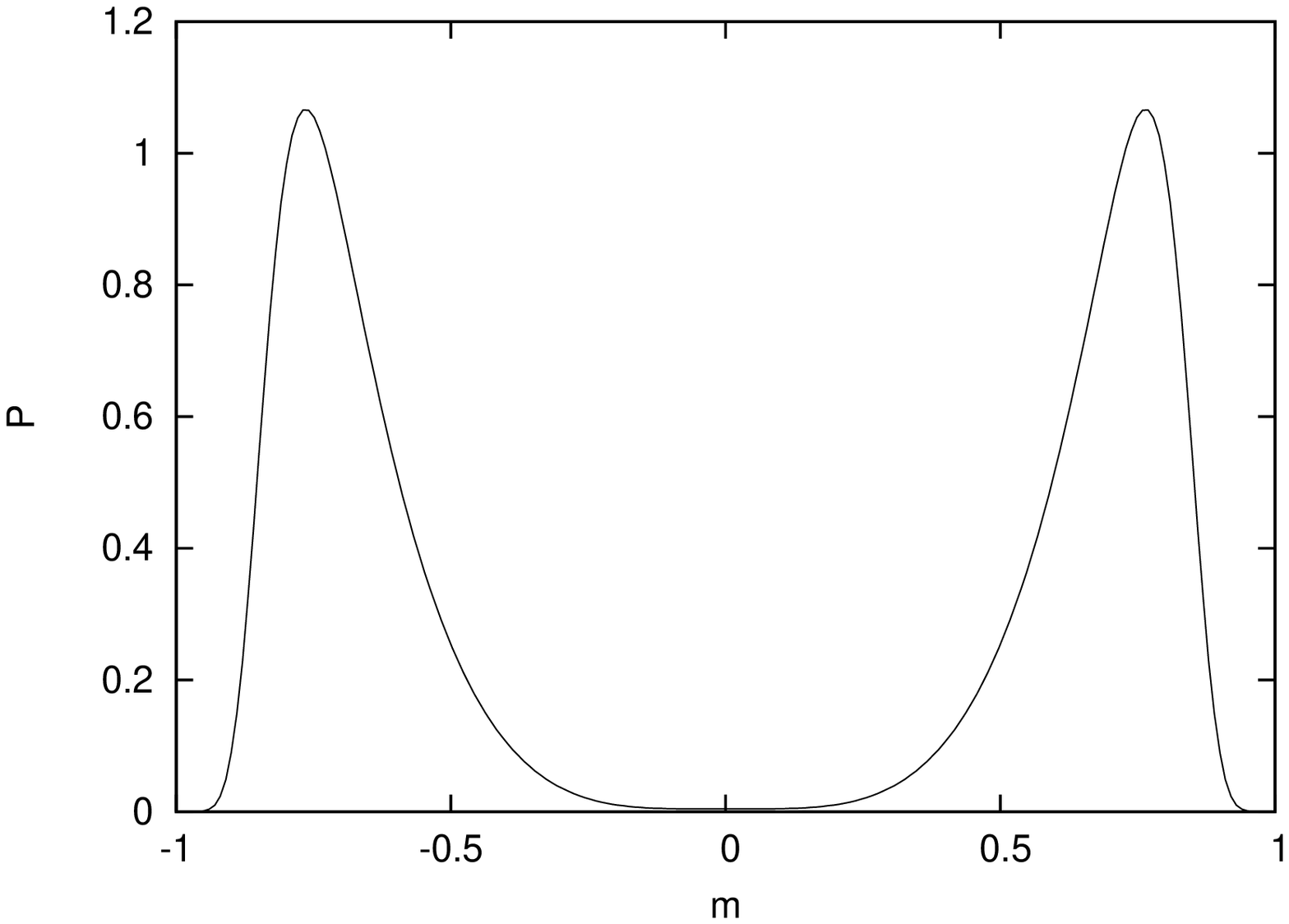}
	\caption{Sketch of the probability distribution $\mathcal{P}$ of the magnetization $m$ for an Ising system in its ferromagnetic phase.}
	\label{fig:MPD}
\end{figure}

This model is known to be in a ferromagnetic phase for large values of the inverse temperature $\beta$. Vaguely speaking, this means that there is a high probability for the system to be in states with nonzero magnetization $m=M/N$ with
\begin{equation}
M(\sigma)=\sum_{i=1}^N\sigma_i,
\end{equation}
but a low probability for observing a magnetization close to zero. Such a probability distribution is given by
\begin{equation}
\mathcal{P}(m)=\mathcal{N}\sum_{\sigma\in S} \delta[M(\sigma)-Nm]\exp[-\beta H(\sigma)],
\end{equation}
where $\mathcal{N}$ is a normalization constant. A sketch of $\mathcal{P}(m)$ might look approximately like the graph shown in Fig.\ \ref{fig:MPD}.

A Monte Carlo estimator of $\mathcal{P}(m)$ is then obtained by computing
\begin{equation}
\widetilde{P}(m)=\frac{1}{N}\sum_{\sigma\in\{X_n\}_{n=0}^{N-1}}\frac{\delta[M(\sigma)-Nm]\exp[-\beta H(\sigma)]}{\pi_\sigma},
\end{equation}
where we decided not to bother about the normalization constant. According to the scheme from Sec.\ \ref{constr}, we have the parameters P2--P4 to choose when constructing the transition matrix of the underlying Markov chain. A standard choice of these parameters is the one also used in the worked example in Sec.\ \ref{constr}:
\begin{description}
\item[P2:] Single spin flip proposal matrix: pick one of the $\sigma_i$ with equal probability $1/N$ and propose to flip it to $-\sigma_i$.
\item[P3:] Stationary distribution $\pi_\sigma\propto\exp[-\beta H(\sigma)]$ (Boltzmann distribution).
\item[P4:] Metropolis acceptance function $f(z)=\min\{1,z\}$.
\end{description}
Knowledge of the approximate shape of $\mathcal{P}(m)$ as sketched in Fig.\ \ref{fig:MPD} indicates already that we might run into trouble with these choices: Imagine a given initial state $\sigma_0$ of the Monte Carlo chain such that its magnetization $M(\sigma_0)$ is located in the left one of the two maxima of the probability distribution $\mathcal{P}(m)$ in Fig.\ \ref{fig:MPD}. The single spin flip proposal P2 will then propose a new state $\sigma_1$ with a magnetization $M(\sigma_1)=M(\sigma_0)\pm 2/N$. For the relatively large numbers of particles $N$ of interest in statistical physics, this corresponds to a small change of magnetization. Therefore, in order to also reach the (numerous and important) states corresponding to the right maximum of $\mathcal{P}$, the process has to proceed step by step through the minimum at magnetization zero. This, however, is exceedingly unlikely to happen, and with high probability the process will remain trapped in the one maximum of $\mathcal{P}$ where it started, inducing a bias in the Monte Carlo estimator.

How can this problem possibly be resolved? Reflecting the way the the transition matrix has been constructed, the strategies employed concentrate either on a modification of the proposal matrix, or on the choice of the stationary distribution (which in turn effects the acceptance matrix). 

\paragraph{Modifying the stationary distribution}
Knowing that the low probability of finding states with magnetizations around zero is causing the problems, one strategy is to choose a stationary distribution of the Markov chain which favors precisely these states. In this way, proposals to move towards zero on the magnetization axis are accepted more frequently, while those moving away from zero are more often rejected. Some examples of this strategy (together with the appropriate references) can be found in the glossary, including entropic sampling, multicanonical ensemble, and umbrella sampling. In order to employ this strategy, a knowledge of the probability distributions of the relevant observables is of avail, and such knowledge may also be acquired by means of preparatory Monte Carlo simulations.

\paragraph{Modifying the proposal matrix}
Instead of favoring the acceptance of states in the vicinity of magnetization zero as in the previous paragraph, an other strategy consists in modifying the proposal matrix such that also states which correspond to a large step on the magnetization axis are suggested. This can be achieved by proposing to flip not only a single spin, but for example a cluster of many adjacent spins. Several variants of such cluster algorithms have been reported in the literature, including the Swendsen-Wang cluster algorithm \citep{SweWa:87} and the Wolff cluster algorithm \citep{Wolff:89}.

Almost all strategies employed in Markov chain Monte Carlo fall into one of these two classes, and although the details may differ, the key ideas are mostly very similar in nature.

\section{Summary}
\label{summary}

In this article, the foundations of Markov-chain Monte Carlo have been presented with the aim of identifying the simulation parameters which still may be adjusted. This formal approach is suitable for exposing the flexibility of the Monte Carlo method, allowing us to draw on plentiful possibilities when designing new and efficient Monte Carlo algorithms.

We have restricted the presentation in this article to Markov-chain Monte Carlo methods with transition matrices that do not depend on time, but discussing a broader class of Monte Carlo methods from the point of view we have advocated is straightforward. Let us mention two important examples:
\begin{description}
\item[Deterministic proposal.---]In the $n$th Monte Carlo step, flipping of the $(n \bmod N)$-th out of $N$ spins is proposed with probability one. This is a time-dependant proposal, resulting in a time-depending transition matrix. The computational advantage of this scheme relies on the fact that per Monte Carlo step one random number less needs to be generated, resulting in a faster algorithm. Convergence issues, however, may be more delicate.
\item[Wang-Landau algorithm.---]This algorithm uses the outcomes of the first $n$ elements of a stochastic process for determining the acceptance rates of the $(n+1)$-th Monte Carlo step, resulting in a non-Markovian stochastic process. This method can significantly speed-up convergence in certain instances \citep{WaLa:01,Landau_etal:04}.
\end{description}
Having studied the material presented in this article, the reader should be able to identify the key ideas of most Monte Carlo algorithms, both Markov chain and beyond. Moreover, it should have become clear how flexible a tool Monte Carlo can be. Of course, the present article covers only certain aspects of Monte Carlo. In order to write a good Monte Carlo simulation, training in the more applied aspects will be essential as well. Hopefully the knowledge of both, the formal and the practical aspects will enable and encourage the reader to give free rein to his creativity as what regards Monte Carlo.

\section*{Acknowledgments}
I am indebted to Michael Promberger who, back at the times when I was a PhD student, shaped my point of view on Monte Carlo techniques. The discussions with him, though long ago, as well as his notes on the topic were of invaluable help when preparing this article.

\begin{appendix}
\section{Proofs}

\begin{proof}[Proof of Theorem \ref{tm_theorem}]\label{tm_proof}
1. ${\mathtt T}$ is a stochastic matrix:
\begin{enumerate}
\renewcommand{\labelenumi}{(\alph{enumi})}
\item ${\mathtt T}_{ij}\geq0$ for all\/ $i,j\in S$: follows immediately from the conditions {\em C\ref{Pstoch}} and {\em C\ref{Apos}} of Theorem \ref{tm_theorem}.
\item The transition probabilities sum up to unity:
\begin{equation}
\begin{split}
\sum_{i\in S}{\mathtt T}_{ij} &= \sum_{i\in S}{\mathtt P}_{ij}{\mathtt A}_{ij}+\sum_{i\in S}\delta_{ij}\sum_{k\in S}{\mathtt P}_{kj}(1-{\mathtt A}_{kj}) = \sum_{i\in S}{\mathtt P}_{ij}{\mathtt A}_{ij}+\sum_{k\in S}{\mathtt P}_{ki}(1-{\mathtt A}_{ki})\\
& = \sum_{i\in S}{\mathtt P}_{ij}{\mathtt A}_{ij}+\sum_{i\in S}{\mathtt P}_{ij}(1-{\mathtt A}_{ij}) = \sum_{i\in S}{\mathtt P}_{ij}\;=\;1,
\end{split}
\end{equation}
where condition {\em C\ref{Pstoch}} has been used in the last line.
\end{enumerate}
2. $\pi$ is the stationary distribution of\/ ${\mathtt T}$:
\begin{equation}
({\mathtt T}\pi)_i  = \sum_{j\in S}{\mathtt T}_{ij}\pi_j  = \sum_{j\in S}{\mathtt P}_{ij}{\mathtt A}_{ij}\pi_j + \sum_{j\in S}\delta_{ij}\sum_{k\in S}{\mathtt P}_{kj}(1-{\mathtt A}_{kj})\pi_j
 = \sum_{j\in S}{\mathtt P}_{ij}{\mathtt A}_{ij}\pi_j + \sum_{k\in S}{\mathtt P}_{ki}(1-{\mathtt A}_{ki})\pi_i.
\end{equation}
Since $\mathtt{P}$ is symmetric we obtain
\begin{equation}
({\mathtt T}\pi)_i = \sum_{j\in S}{\mathtt P}_{ij}({\mathtt A}_{ij}\pi_j-{\mathtt A}_{ji}\pi_i) + \pi_i\sum_{j\in S}{\mathtt P}_{ij},
\end{equation}
which yields $\pi_i$ as a consequence of detailed balance.
\end{proof}

\begin{proof}[Proof of Theorem \ref{is_theorem}]\label{is_proof}

The variance of $\widetilde{F}$ is defined as
\begin{equation}\label{var_def}
\frac{1}{N}\sum_{i\in\{X_n\}_{n=1}^N}\biggl(\frac{f(i)}{\pi_i}\biggr)^2-\biggl(\frac{1}{N}\sum_{i\in\{X_n\}_{n=1}^N}\frac{f(i)}{\pi_i}\biggr)^2.
\end{equation}
Since $\{X_n\}_{n=1}^N$ is distributed according to $\pi$, the second term is independent of $\pi$. We therefore only need to minimize the first term, for which Jensen's inequality provides the lower bound
\begin{equation}
\frac{1}{N}\sum_{i\in\{X_n\}_{n=1}^N}\left(\frac{f(i)}{\pi_i}\right)^2 \geqslant\frac{1}{N^2}\biggl(\sum_{i\in\{X_n\}_{n=1}^N}\frac{|f(i)|}{\pi_i}\biggr)^2=\biggl(\sum_{i\in S}|f(i)|\biggr)^2.
\end{equation}
Inserting $\displaystyle\pi_i=|f(i)|/\sum_{j\in S}|f(j)|$ into the left hand side of this inequality, the lower bound is attained,
\begin{equation}
\frac{1}{N}\sum_{i\in\{X_n\}_{n=1}^N}\biggl(\frac{f(i)}{\pi_i}\biggr)^2=\frac{1}{N}\sum_{i\in\{X_n\}_{n=1}^N}\biggl(\sum_{j\in S}|f(j)|\biggr)^2=\biggl(\sum_{i\in S}|f(i)|\biggr)^2.\qedhere
\end{equation}
\end{proof}

\end{appendix}

\section*{Glossary}\label{glossary}
\addcontentsline{toc}{section}{Glossary}
\setcounter{equation}{0}
\renewcommand{\theequation}{G\arabic{equation}}

In this glossary, a few of the most widely used methods and strategies in statistical physics Monte Carlo are formulated in the language introduced in the main body of the article. In this way, the structure of a particular algorithm and the main idea behind it are exposed particularly clearly, which makes for example the $\rightarrow${\em Creutz demon}\/ appear a bit less demonic than in its original formulation. The list is far from being complete, but it should be useful for bringing to life the general ideas outlined in this article.




To be explicit, whenever any of the concepts in this glossary need to be described in terms of a state space function, the Hamiltonian function $H :S\to\rz$ is chosen, and the value of the Hamiltonian function for a certain state is called energy $E$. A generalization to other functions or more than one function is possible and straightforward. 


\begin{description}

\item[Boltzmann distribution.---]Distribution $\pi_i\propto\exp[-\beta H (i)]$ where $\beta$ is a parameter (inverse temperature).


\item[Canonical distribution.---]$\rightarrow${\em Boltzmann distribution}.

\item[Cluster algorithm.---]Proposal matrix which, for a spin system, in order to improve the performance of a simulation (see Sec.\ \ref{speed}), proposes to change the value not of a single spin variable ($\rightarrow${\em Single spin flip algorithm}), but of a cluster of spin variables at a time. A cluster is a connected collection of spin variables of the same value. 
Examples include the Swendsen-Wang cluster algorithm \citep{SweWa:87} and the Wolff cluster algorithm \citep{Wolff:89}.

\item[Creutz demon.---]Step shaped stationary distribution $\pi_i\propto\Theta(E- H (i))$, where $E$ is a parameter. The same name is also used for the distribution $\pi_i\propto\Theta(-E_1+ H (i))\Theta(E_2- H (i))$ where $E_1<E_2$. 


\item[Dynamical ensemble.---]Power law shaped stationary distribution $\pi_i\propto \left[(E- H (i))/B\right]^{(B-2)/2}$, where $E$ and $B$ are parameters \citep{HueGe}.

\item[Entropic sampling.---]$\rightarrow${\em Umbrella sampling}.



\item[Gaussian ensemble.---]Stationary distribution $\pi_i\propto\exp\left\{-a[ H(i)-E]^2\right\}$, where $a$ and $E$ are parameters. In the limit $a\to\infty$, a $\rightarrow${\em Microcanonical distribution} is approached \citep{Hetherington}.

\item[Glauber dynamics.---]Acceptance function of the form $f:z\mapsto z/(1+z)$ \citep{Glauber}.


\item[Histogram.---]For a Markov chain $\{X_n\}_{n=1}^N$, a histogram is the number of states which are subject to a certain condition. Often energy histograms are employed, i.e., the number of states
\begin{equation}
h(E):=\sum_{i\in\{X_n\}_{n=1}^N}\delta_{E, H (i)}
\end{equation}
from a Markov chain with a certain energy $E$. For the density of states
\begin{equation}
\Omega(E)=\sum_{i\in S}\delta_{E,H(i)}
\end{equation}
of a system, an estimator $\widetilde{\Omega}=h(E)/(N\pi(E))$ can be obtained from a histogram, where $\pi(E)$ is the (energy dependent) stationary distribution of the Markov chain. First applied in statistical physics Monte Carlo by \citet{SaJaFiWo}.

\item[Histogram reweighting.---]Essentially a calculation of canonical quantities from their definition: an estimator of a canonical average is computed from a Monte Carlo estimator of the density of states (which in turn is obtained from a $\rightarrow${\em Histogram}). First applied in connection with Monte Carlo by \citet{SaJaFiWo}

\item[Kawasaki dynamics.---]Proposal matrix which, for a spin system, restricts the Markov chain onto a subspace of the state space such that the magnetization is kept fixed \citep{Kawasaki}. This is achieved by proposing to interchange the values of the spins at two randomly chosen sites (or at two randomly chosen neighboring sites). 

\item[Metropolis acceptance.---]Popular choice of the acceptance function, $f:z\mapsto\min\{1,z\}$ \citep{MeRoRoTeTe}.


\item[Microcanonical distribution.---]Distribution $\pi_i\propto\delta[E-H (i)]$.


\item[Multicanonical ensemble.---]$\rightarrow${\em Umbrella sampling}.

\item[Multihistogram method.---]Combining the information contained in $\rightarrow${\em His\-to\-grams} from two or more distinct Markov chains, typically for different transition matrices, in a way such that statistical errors in the resulting data are minimized. For two histograms, $h_1 (E)$ and $h_2 (E)$, assuming the statistical errors $\Delta h_i (E)=\sqrt{h_i (E)}$ to be those of a Poisson distribution, the statistical error in the density of states $\Omega(E)$ is minimized for
\begin{equation}
\Omega(E)=\frac{h_1 h_2[\pi_1+a\pi_2]}{h_1\pi_2^2+h_2\pi_1^2},
\end{equation}
where
\begin{equation}
a=\sum\limits_E \frac{h_1\pi_1h_2\pi_2}{h_1\pi_2^2+h_2\pi_1^2} \Big/ \sum\limits_E\frac{h_2^2\pi_1^2}{h_1\pi_2^2+h_2\pi_1^2}.
\end{equation}
Put forward by \citet{Bennett}, extended by \citet{FeSwe2}. 


\item[Parallel tempering.---]$\rightarrow${\em Replica exchange Monte Carlo}.

\item[Replica exchange Monte Carlo.---]As in the related method of $\rightarrow${\em Simulated tempering}, improvement of convergence is achieved by enlarging the state space of the Markov chain: for a quantity of interest which consists of a sum over a state space $S$,
\begin{equation}
F=\sum_{i\in S}f(i),
\end{equation}
an estimator $\widetilde{F}$ is obtained by making use of a Markov chain not on $S$, but on a superspace of $S$,
\begin{equation}\label{partemp}
\widetilde{F}_{\lambda,{\mathtt T},N,\widetilde{S}}=\frac{1}{N}\sum\frac{1}{|\widetilde{S}|}\sum_{j=1}^{|\widetilde{S}|}\frac{f(i_j)}{\pi_{i_j,m_j}},
\end{equation}
where the first sum is over all indices $(i_1,m_1,\dotsc,i_{|\widetilde{S}|},m_{|\widetilde{S}|})\in\{X_n\}_{n=1}^N$. In contrast to simulated tempering, the chain $\{X_n\}_{n=1}^N$ is $\mbox{Markov}(\lambda,{\mathtt T},N)$ on an even larger space, the product space $(S\times\widetilde{S})^{|\widetilde{S}|}$ (where $|\widetilde{S}|$ denotes the number of elements in $\widetilde{S}$), restricted such that, for each element $(i_1,m_1,\dotsc, i_{|\widetilde{S}|},m_{|\widetilde{S}|})$ from this state space, the $m_j$ form a permutation of the elements of $\widetilde{S}$. The stationary distribution on the product space is again a product, $(\pi_{i,m})^{|\widetilde{S}|}$. The method can be viewed as creating simultaneously stochastic chains on $|\widetilde{S}|$ copies of the same system but with different stationary distributions, while exchanging randomly the stationary distributions of the copies. The exchange of the distributions mirrors the above ``permutation''-restriction of the state space. The expectation values from these copies can be summed up to improve the Monte Carlo estimator [second sum in Eq.\ (\ref{partemp})]. The advantage with respect to simulated tempering is that, due to the exchange of distributions, the different $m$-values appear equally distributed and no laborious tuning of additional constants is necessary. Put forward by \citet{SweWa:86}.

\item[Simulated tempering --]
This method aims at improving convergence by using a generalization of definition \ref{def_mcestimator} of a Monte Carlo estimator given in Sec.~\ref{mcsim}: for a quantity of interest which consists of a sum over a state space $S$,
\begin{equation}
F=\sum_{i\in S}f(i),
\end{equation}
an estimator $\widetilde{F}$ is obtained by making use of a Markov chain not on $S$, but on a superspace of $S$,
\begin{equation}
\widetilde{F}_{\lambda,{\mathtt T},N,\widetilde{S}} =\frac{1}{N}\sum_{(i,m)\in\{X_n\}_{n=1}^N}\frac{f(i)}{\pi_{i,m}}
\end{equation}
where $\{X_n\}_{n=1}^N$ is $\mbox{Markov}(\lambda,{\mathtt T},N)$ on the product space $S\times\widetilde{S}$. A careful choice of $\widetilde{S}$ and the stationary distribution $\pi_{i,m}$ can lead to a significant improvement of convergence. In the original paper by \citet{MaPa}, $\pi_{i,m}$ is chosen Boltzmann distributed $\exp[-\beta_m  H (i)+g_m]$ with different inverse temperatures $\beta_m$ and constants $g_m$, labeled by the indices $m\in\widetilde{S}$. Then, switching between states with different $m$-values can be regarded as considering a system at randomly switching temperatures. Careful tuning of the $g_m$-values is necessary in order to obtain non-negligible probabilities for having states of the various $m$-values in the Markov chain.


\item[Single spin flip algorithm.---]Proposal matrix which, for a spin system, proposes to change the value of only one of the spin variables at a time. This guarantees proposal of a state with an energy similar to that of the current state, and therefore a high acceptance rate (see Sec.\ \ref{speed}). A more general formulation is the following: on a state space $S$ with product structure, $S=\Sigma\times...\times\Sigma=\Sigma^M$, and states $i=(i_1,...,i_M)\in S$ where $i_m\in\Sigma$ for all $m=1,...,M$, a single spin flip algorithm employs a proposal matrix ${\mathtt P}=({\mathtt P}_{ij}:i,j\in S)$ with elements
\begin{equation}
{\mathtt P}_{ij}=\begin{cases}
\frac{1}{M(|\Sigma|-1)} & \text{if $\sum_{m=1}^M \delta_{i_m,j_m}=M-1$},\\
0 & \text{else},
\end{cases}
\end{equation}
where $|\Sigma|$ is the number of elements in $\Sigma$.

\item[Umbrella sampling.---]Stationary distribution $\displaystyle \pi_i\propto 1 / \widetilde{\Omega}[ H (i)]$ where $\widetilde{\Omega}$ is an estimator on the density of states $\Omega(E):=|S|^{-1}\sum_{i\in S}\delta_{E, H (i)}$. This choice of the stationary distribution gives rise to an approximately flat energy $\rightarrow${\em Histogram},
\begin{equation}
h(E)=\frac{1}{N}\sum_{i\in\{X_n\}_{n=1}^N}\delta_{E, H (i)}\propto\frac{1}{N}\sum_{i\in S}\frac{\delta_{E, H (i)}}{\widetilde{\Omega}(E)}\approx\text{const.}
\end{equation}
Put forward by \citet{ToVa}, but variations of this idea reappeared in the literature under the names of multicanonical ensemble \citep{BeNe} and entropic sampling \citep{Lee}.


\end{description}

\bibliographystyle{elsarticle-harv.bst}
\bibliography{simrev.bib}

\begin{thebibliography}{27}
\expandafter\ifx\csname natexlab\endcsname\relax\def\natexlab#1{#1}\fi
\expandafter\ifx\csname url\endcsname\relax
  \def\url#1{\texttt{#1}}\fi
\expandafter\ifx\csname urlprefix\endcsname\relax\def\urlprefix{URL }\fi

\bibitem[{Bauer(1995)}]{Bauer}
Bauer, H., 1995. Probability Theory. De Gruyter Studies in Mathematics. De
  Gruyter, Berlin.

\bibitem[{Bennett(1976)}]{Bennett}
Bennett, C.~H., 1976. Efficient estimation of free energy differences from
  {M}onte {C}arlo data. J. Comput. Phys. 22, 245--268.

\bibitem[{Berg and Neuhaus(1991)}]{BeNe}
Berg, B.~A., Neuhaus, T., 1991. Multicanonical algorithms for first order phase
  transitions. Phys. Lett. B 267, 249--253.

\bibitem[{Boghosian(1999)}]{Boghosian}
Boghosian, B.~M., 1999. A generalization of {M}etropolis and heat-bath sampling
  for {M}onte {C}arlo simulations. Phys. Rev. E 60, 1189--1194.

\bibitem[{Caracciolo et~al.(1994)Caracciolo, Pelissetto, and Sokal}]{CaPeSo}
Caracciolo, S., Pelissetto, A., Sokal, A.~D., 1994. A general limitation on
  {M}onte {C}arlo algorithms of {M}etropolis type. Phys. Rev. Lett. 72,
  179--182.

\bibitem[{Ferrenberg and Swendsen(1989)}]{FeSwe2}
Ferrenberg, A.~M., Swendsen, R.~H., 1989. Optimized {M}onte {C}arlo data
  analysis. Phys. Rev. Lett. 63, 1195--1198.

\bibitem[{Glauber(1963)}]{Glauber}
Glauber, R.~J., 1963. Time-dependent statistics of the {I}sing model. J. Math.
  Phys. 4, 294--307.

\bibitem[{Hastings(1970)}]{Hastings}
Hastings, W.~K., 1970. Monte {C}arlo sampling methods using {M}arkov chains and
  their applications. Biometrika 57, 97--109.

\bibitem[{Hetherington(1987)}]{Hetherington}
Hetherington, J.~H., 1987. Solid\/ $^3${H}e magnetism in the classical
  approximation. J. Low Temp. Phys. 66, 145--154.

\bibitem[{H\"uller and Gerling(1993)}]{HueGe}
H\"uller, A., Gerling, R.~W., 1993. First order phase transitions studied in
  the dynamical ensemble: The $q$-states {P}otts model as a test case. Z. Phys.
  B 90, 207--214.

\bibitem[{Kawasaki(1966)}]{Kawasaki}
Kawasaki, K., 1966. Diffusion constants near the critical point for
  time-dependent {I}sing models. {I}. Phys. Rev. 145, 224--230.

\bibitem[{Landau and Binder(2000)}]{LandauBinder}
Landau, D.~P., Binder, K., 2000. A Guide to Monte Carlo Simulations in
  Statistical Physics. Cambridge University Press, Cambridge.

\bibitem[{Landau et~al.(2004)Landau, Tsai, and Exler}]{Landau_etal:04}
Landau, D.~P., Tsai, S.-H., Exler, M., 2004. A new approach to {M}onte {C}arlo
  simulations in statistical physics: {W}ang-{L}andau sampling. Am. J. Phys.
  72, 1294--1302.

\bibitem[{Lee(1993)}]{Lee}
Lee, J., 1993. New {M}onte {C}arlo algorithm: entropic sampling. Phys. Rev.
  Lett. 71, 211--214.

\bibitem[{Marinari and Parisi(1992)}]{MaPa}
Marinari, E., Parisi, G., 1992. Simulated tempering: a new {M}onte {C}arlo
  scheme. Europhys. Lett. 19, 451--458.

\bibitem[{Metropolis et~al.(1953)Metropolis, Rosenbluth, Rosenbluth, Teller,
  and Teller}]{MeRoRoTeTe}
Metropolis, N., Rosenbluth, A.~W., Rosenbluth, M.~N., Teller, A.~H., Teller,
  E., 1953. Equation of state calculations by fast computing machines. J. Chem.
  Phys. 21, 1087--1092.

\bibitem[{Newman and Barkema(1999)}]{NewmanBarkema}
Newman, M. E.~J., Barkema, G.~T., 1999. Monte Carlo Methods in Statistical
  Physics. Oxford University Press, Oxford.

\bibitem[{Norris(1998)}]{Norris}
Norris, J.~R., 1998. Markov Chains. Cambridge University Press, Cambridge.

\bibitem[{Rubinstein(1981)}]{Rubinstein}
Rubinstein, R.~Y., 1981. Simulation and the Monte Carlo Method. J. Wiley, New
  York.

\bibitem[{Salsburg et~al.(1959)Salsburg, Jacobson, Fickett, and
  Wood}]{SaJaFiWo}
Salsburg, Z.~W., Jacobson, J.~D., Fickett, W., Wood, W.~W., 1959. Application
  of the {M}onte {C}arlo method to the lattice gas model. {I}.
  {T}wo-dimensional triangular lattice. J. Chem. Phys. 30, 65--72.

\bibitem[{Swendsen(1979)}]{Swendsen:79}
Swendsen, R.~H., 1979. Monte {C}arlo renormalization group. Phys. Rev. Lett.
  42, 859--861.

\bibitem[{Swendsen and Wang(1986)}]{SweWa:86}
Swendsen, R.~H., Wang, J.-S., Nov 1986. Replica {M}onte {C}arlo simulation of
  spin-glasses. Phys. Rev. Lett. 57~(21), 2607--2609.

\bibitem[{Swendsen and Wang(1987)}]{SweWa:87}
Swendsen, R.~H., Wang, J.-S., 1987. Nonuniversal critical dynamics in {M}onte
  {C}arlo simulations. Phys. Rev. Lett. 58, 86--88.

\bibitem[{Torrie and Valleau(1977)}]{ToVa}
Torrie, G.~M., Valleau, J.~P., 1977. Nonphysical sampling distributions in
  {M}onte {C}arlo free-energy estimation: {U}mbrella sampling. J. Comput. Phys.
  23, 187--199.

\bibitem[{Wang and Landau(2001)}]{WaLa:01}
Wang, F., Landau, D.~P., 2001. Efficient, multiple-range random walk algorithm
  to calculate the density of states. Phys. Rev. Lett. 86, 2050--2053.

\bibitem[{Wang and Swendsen(2002)}]{WaSwe}
Wang, J.-S., Swendsen, R.~H., 2002. Transition matrix {M}onte {C}arlo method.
  J. Stat. Phys. 106, 245--285.

\bibitem[{Wolff(1989)}]{Wolff:89}
Wolff, U., 1989. Collective {M}onte {C}arlo updating for spin systems. Phys.
  Rev. Lett. 62, 361--364.

\end{thebibliography}


\end{document}